\documentclass[12pt]{iopart}

\usepackage{
	amssymb, 
	amsthm, 
	graphicx, 
	xspace,
	overpic,
	ifthen,
	verbatim, 
	subfigure, 
	appendix,
	ifthen,
	nicefrac, 
	color, 
	tensor, 
	etoolbox, 
	centernot, 
} 

\usepackage[normalem]{ulem}

\newcommand{\ket}[1]{\vert #1 \rangle  }

\newcommand{\braket}[2]{  \langle #1 \vert #2 \rangle  }

\newcommand{\matrixelement}[3]{  \langle #1 \vert #2 \vert #3\rangle  }

\newcommand{\projector}[1]{  \vert #1 \rangle \langle #1 \vert }

\newcommand{\abs}[1]{| #1 |} 
\newcommand{\Abs}[1]{\left| #1 \right|}
\newcommand{\avg}[1]{\langle #1 \rangle} 

\newcommand{\aavg}[1]{\llangle #1 \rrangle} 

\newcommand{\MI}{\ensuremath{\mathcal{I}}}
\newcommand{\Sys}{\ensuremath{\mathcal{S}}}
\newcommand{\Env}{\ensuremath{\mathcal{E}}}
\newcommand{\Frag}{\ensuremath{\mathcal{F}}}
\newcommand{\FragBar}{{\ensuremath{{\overline{\mathcal{F}}}}}}

\newcommand{\Glob}{\ensuremath{\mathcal{H}}}
\newcommand{\sharpof}[1]{\ensuremath{\tensor[^\sharp]{#1}{}}}

\newcommand{\draftmode}{1}
\newcommand{\oldtext}{\ifnum \draftmode=1 \color[gray]{0.5} \else \color{black} \fi}
\newcommand{\notetoself}[1]{\ifnum \draftmode=1 {\color[rgb]{0,0,0.8} [#1]} \fi}
\newcommand{\cuttext}[1]{\ifnum \draftmode=1 {\color[rgb]{0,0.5,0} [#1]} \fi}

\usepackage{setstack,bm}

\newcommand{\eqref}[1]{(\ref{#1})}

\newcommand{\arctanh}{\; \mathrm{arctanh} \;}
\newcommand{\llangle}{\langle \! \langle}
\newcommand{\rrangle}{\rangle \! \rangle}

\pdfoutput=1
\newcommand{\pbwidthfactor}{0.96}
\newcommand{\medfactor}{0.48}

\renewcommand{\draftmode}{0} 

\begin{document}


\title{The rise and fall of redundancy in decoherence and quantum Darwinism}
\date{\today}
\author{C~Jess~Riedel$^{1,2}$, Wojciech~H~Zurek$^{1,3}$, and Michael Zwolak$^{1,4}$}
\address{$^1$Theoretical Division/CNLS, LANL, Los Alamos, New Mexico 87545, USA}
\address{$^2$Department of Physics, University of California, Santa Barbara, California 93106, USA}
\address{$^3$Santa Fe Institute, Santa Fe, New Mexico 87501, USA}
\address{$^4$Department of Physics, Oregon State University, Corvallis, OR 97331, USA}
\eads{criedel@physics.ucsb.edu}

\pacs{03.67.-a, 03.67.Bg, 03.65.Yz}

\begin{abstract}
A state selected at random from the Hilbert space of a many-body system is overwhelmingly likely to exhibit highly non-classical correlations.  For these typical states, half of the environment must be measured by an observer to determine the state of a given subsystem.  The objectivity of classical reality---the fact that multiple observers can agree on the state of a subsystem after measuring just a small fraction of its environment---implies that the correlations found in nature between macroscopic systems and their environments are very exceptional.  Building on previous studies of quantum Darwinism showing that highly redundant branching states are produced ubiquitously during pure decoherence, we examine conditions needed for the creation of branching states and study their demise through many-body interactions.  We show that even constrained dynamics can suppress redundancy to the values typical of random states on relaxation timescales, and prove that these results hold exactly in the thermodynamic limit.
\end{abstract}
\maketitle


Hilbert space is a big place,  exponentially larger than the arena of classical physics.  The Hilbert space of macroscopic systems is dominated by states that have no classical counterparts.  Yet the world observed by macroscopic observers exhibits powerful regularities that make it amenable to classical interpretations on a broad range of scales.  How do we explain this?

The answer, of course, is that Hilbert space is not sampled uniformly; rather, the initial state and the Hamiltonian governing evolution are both very special.  Quantum Darwinism \cite{Zurek2000, Zurek2009} is a framework for describing and quantifying what distinguishes quasi-classical states awash in the enormous sea of Hilbert space. 

Typical macroscopic observers do not directly interact with a system.  Instead, they sample a (small) part of its environment in order to infer its state, using the environment as an information channel \cite{Ollivier2004}.  Thus, when we measure the position of a chair by looking at it, our eyes do not directly interact with the chair. By opening our eyes, we merely allow them (and hence, our neurons) to become correlated with some of the photons scattered by chair (and hence, its position).  

Consider a system $\Sys$ with Hilbert space of dimension $D_\Sys$  decohered by a multi-partite environment $\Env = \bigotimes_{i=1}^N \Env_i$, where each $\Env_i$ has dimension $D_\Env$.  To understand the perception of classicality by macroscopic observers, it is of great interest to understand the quantum mutual information between $\Sys$ and some subset of the environment (a \emph{fragment}) $\Frag =  \bigotimes_{i \in F} \Env_i$, where $F \subset \{1, \ldots, N\}$:
\begin{eqnarray}
\MI_{\Sys : \Frag} &= H_{\Sys} + H_{\Frag} - H_{\Sys \Frag} .
\end{eqnarray}
Above, $H_{\Sys}$, $H_{\Frag}$, and $H_{\Sys \Frag}$ are the respective individual and joint von Neumann entropies.  We denote the size of the fragment by $\sharpof{\Frag} = \Abs{F} = fN$, where $f \equiv \sharpof{\Frag}/N$ is the fraction of $\Env$ contained in $\Frag$.  The mutual information averaged over all $\Frag$ of a given fractional size $f$ is written as
\begin{eqnarray}
\bar{\MI}(f) = \avg{\MI_{\Sys : \Frag}}_{\sharpof{\Frag}} .
\end{eqnarray}
When the global state of $\Sys \Env$ is pure, one can show \cite{Blume-Kohout2005} that this function is non-decreasing and anti-symmetric about its value at $f = \nicefrac{1}{2}$.   

In the absence of preferred initial states or dynamics, the natural question is: what is the typical amount of mutual information between $\Sys$ and $\Frag$, and how does it depend on the fractional size $f$ of the fragment?  To be quantitative, we use the Haar measure on the space of pure states in the global Hilbert space $\Glob = \Sys \otimes \Env$ of dimension $D = D_\Sys D_\Env^N$.  (This is the natural, unique unitarily invariant measure on this space.)  Page's formula for the Haar-average entropy of a subsystem \cite{Page1993, Sanchez-Ruiz1995, Foong1994} can be used to calculate \cite{Blume-Kohout2005} the average of $\bar{\MI}$ over $\Glob$.  
If we hold $f$ fixed, we find that $\lim_{N \to \infty} \avg{\bar{\MI} (f)}_\Glob = 0$ if $f< \nicefrac{1}{2}$. In other words, for a randomly selected pure state in the global Hilbert space, an observer typically cannot learn \emph{anything} about a system without sampling at least \emph{half} its environment.  States that deviate (even by exponentially small amounts) from this property occupy an exponentially small volume in Hilbert space \cite{Hayden2004} as $N \to \infty$.  (This is a consequence of the mathematical phenomenon known as the ``concentration of measure'' in high-dimensional spaces \cite{Ledoux2001}, which can be thought of as an abstract law of large numbers.)

It's natural to define the \emph{redundancy} $R_{\delta}$ as the number of distinct fragments in the environment that supply, up to an information deficit $\delta$, the classical information about the state of the system.  More precisely, $R_{\delta} = 1/f_\delta$, where $f_\delta$ is the smallest fragment such that $\bar{\MI}(f_\delta) \ge (1-\delta)H_\Sys^\mathrm{max}$, and $H_\Sys^\mathrm{max}$ is the maximum entropy of $\Sys$.  The dependence on $\delta$ is typically \cite{Zwolak00} only logarithmic.  At any given time, the redundancy is the measure of objectivity; it counts the number of observers who could each independently determine the approximate state of the system by interacting with disjoint fragments of the environment.  As described in the previous paragraph, typical states in $\Glob$ will have $\bar{\MI} (f) \approx 0$ for $f < \nicefrac{1}{2}$ and, by symmetry, $\bar{\MI} (f) \approx 2 H_\Sys^{\mathrm{max}}$ for $f > \nicefrac{1}{2}$ , so $R_\delta \approx 2$ for any $\delta$.  That is, half the environment must be captured to learn anything about $\Sys$.  These states are essentially \emph{non-redundant}, and make up the vast bulk of Hilbert space. 


\begin{figure} [bt]
 \centering 
  \includegraphics[width=\medfactor\columnwidth]{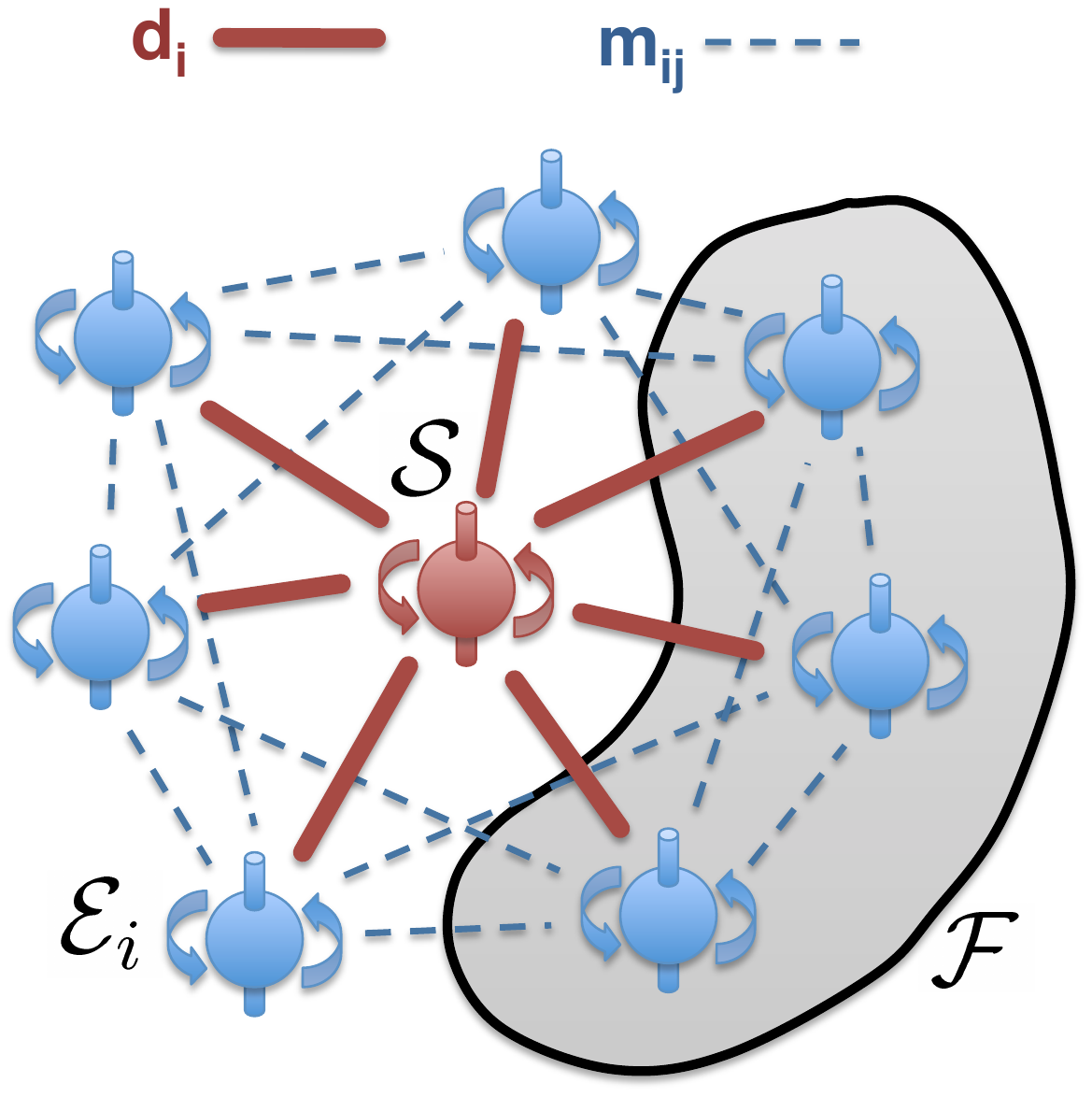}  
  \caption{We investigate an environment of 16 spins $\Env_i$ coupled to a single system qubit $\Sys$ with Hamiltonian and initial state given by Eqs.~\eqref{hamiltonian} and \eqref{initial-state}.  A fragment $\Frag$ is a subset of the whole environment $\Env$.  The couplings $d_i$ and $m_{i j}$ were selected from normal distributions with zero mean and standard deviations $\sigma_d = 0.1$ and $\sigma_m = 0.001$.  Crucially, the interactions between $\Sys$ and the $\Env_i$ are much stronger than those within $\Env$.  That is, $\sigma_d \gg \sigma_m$.
}
   \label{diagram}
\end{figure}

\begin{figure} [b!]
  \centering 
   \includegraphics[width=\pbwidthfactor\textwidth]{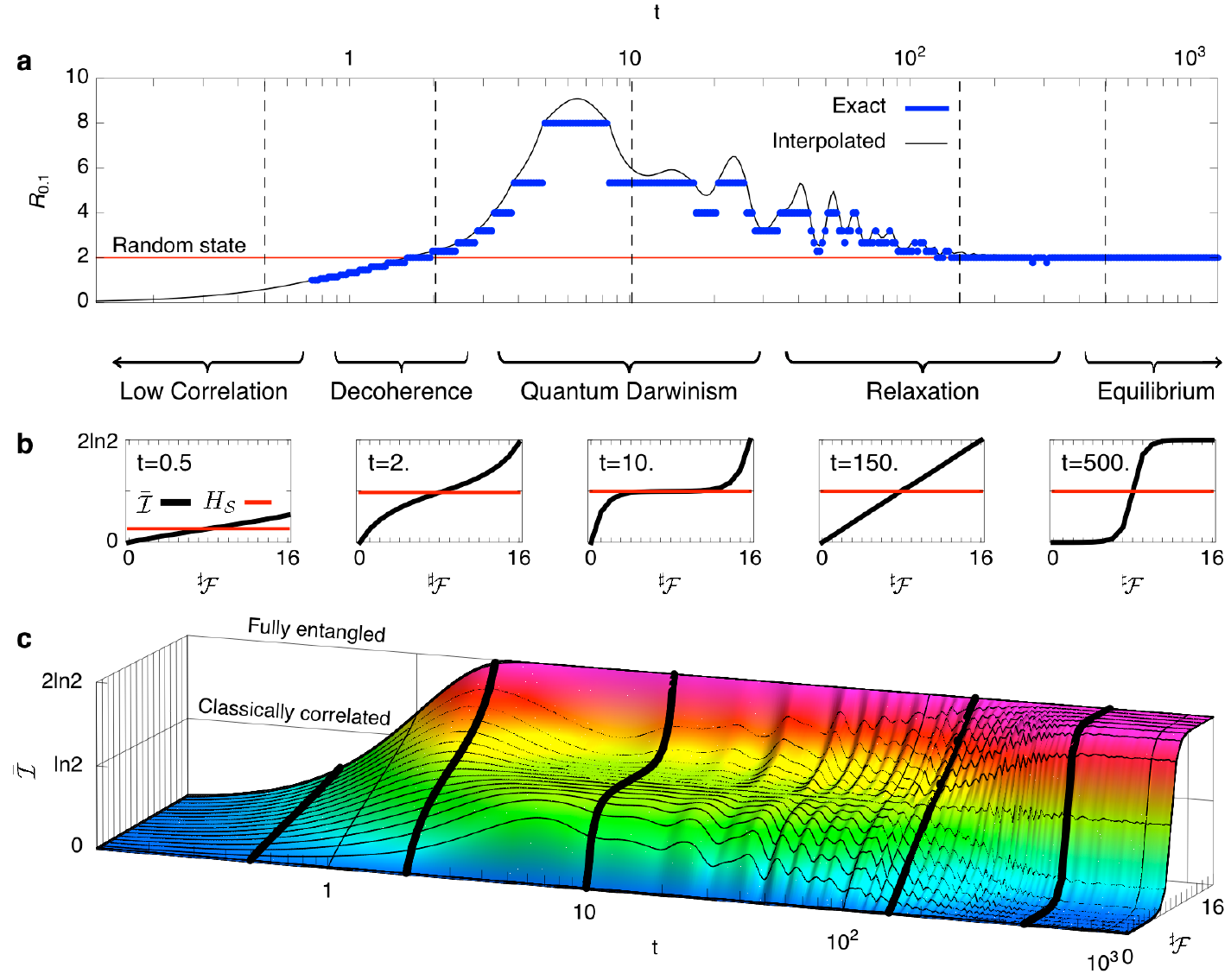} 
       \caption{(Caption next page.)}
\end{figure}
\addtocounter{figure}{-1}
\begin{figure} [t!]
  \caption{(Previous page.)We study the spin universe described in figure \ref{diagram}.
(a) The redundancy $R_\delta$ is the number of fragments of $\Env$ that provide, up to a fractional deficit $\delta = 0.1$, complete classical information about the system.   The exact redundancy is supplemented by an estimate based on the linearly interpolated value of $\bar{\MI} (f)$ to guide the eye.  This can be compared to $R_\delta \approx 2$, the redundancy of nearly all states in the global Hilbert space.   The vertical dashed lines mark five time slices. 
(b) The mutual information $\bar{\MI}$ versus fragment size $\sharpof{\Frag} $, and the entropy $H_\Sys$ of the system, at five time slices corresponding to different qualitative behavior.
(c) The complete mutual information $\bar{\MI}$ versus both fragment size $\sharpof{\Frag} $ and time $t$.  The five time slices are marked with thick black lines. 
\textbf{Low correlation} ($t=0.5$) for small times means the environment ``knows'' very little about the system.  Each spin added to $\Frag$ reveals a bit more about $\Sys$, resulting in the linear dependence of $\bar{\MI}$.   
\textbf{Decoherence} ($t=2$) sets in near $\tau_d \equiv (\sqrt{N} \sigma_d)^{-1} = 2.5$. By that time, the density matrix of $\Sys$ is approximately a mixture of the two pointer states $\ket{\uparrow}$ and $\ket{\downarrow}$ singled out by the interaction Hamiltonian.   Mutual information is still nearly linear in $\sharpof{\Frag}$ and redundancy is of order unity. Mixing within the environment can be neglected because $t \ll \sigma_m^{-1} = 1000$.  
\textbf{Quantum Darwinism} ($t=10$) is characterized by a mutual information plot that rises quickly to the classical plateau; the first few spins in a fragment give essentially all classical information, and additional spins just confirm what is already known. The remaining quantum information (above the plateau) is still present in the global state but it is effectively inaccessible, in that it can only be recovered by an unrealistic observer accurately measuring the joint state of almost all of $\Env$.  After $t \sim \sigma_d^{-1} = 10$, only order unity spins are needed to determine the state of $\Sys$ no matter how large $N$ is, so $R_\delta \sim N$.  In the absence of the couplings $m_{i j}$  between environment fragments this situation would persist forever.  (For some environments, such as photons, this is indeed the case.)  
\textbf{Relaxation} ($t=150$) occurs near $t \sim \tau_m \equiv (\sqrt{N} \sigma_m)^{-1}  = 250$.  Mixing within the environment entangles any given fragment's information about the system with the rest of the environment, reducing the usefulness of measurements on that fragment.  The mutual information plateau is destroyed, so redundancy plummets.  
\textbf{Equilibrium} ($t=500$) is reached for $t \sim \sigma_m^{-1} = 1000$, when the actions associated with interaction between individual spin pairs in the environment reach order unity.   The mutual information plot takes the non-redundant form characteristic of a random states in the combined Hilbert space of $\Sys \Env$.  An observer can learn nothing about the system unless he samples almost half the environment.}
\label{spins-PIP}

\end{figure}


\begin{figure} [b!]
  \centering 
  \includegraphics[width=\pbwidthfactor\textwidth]{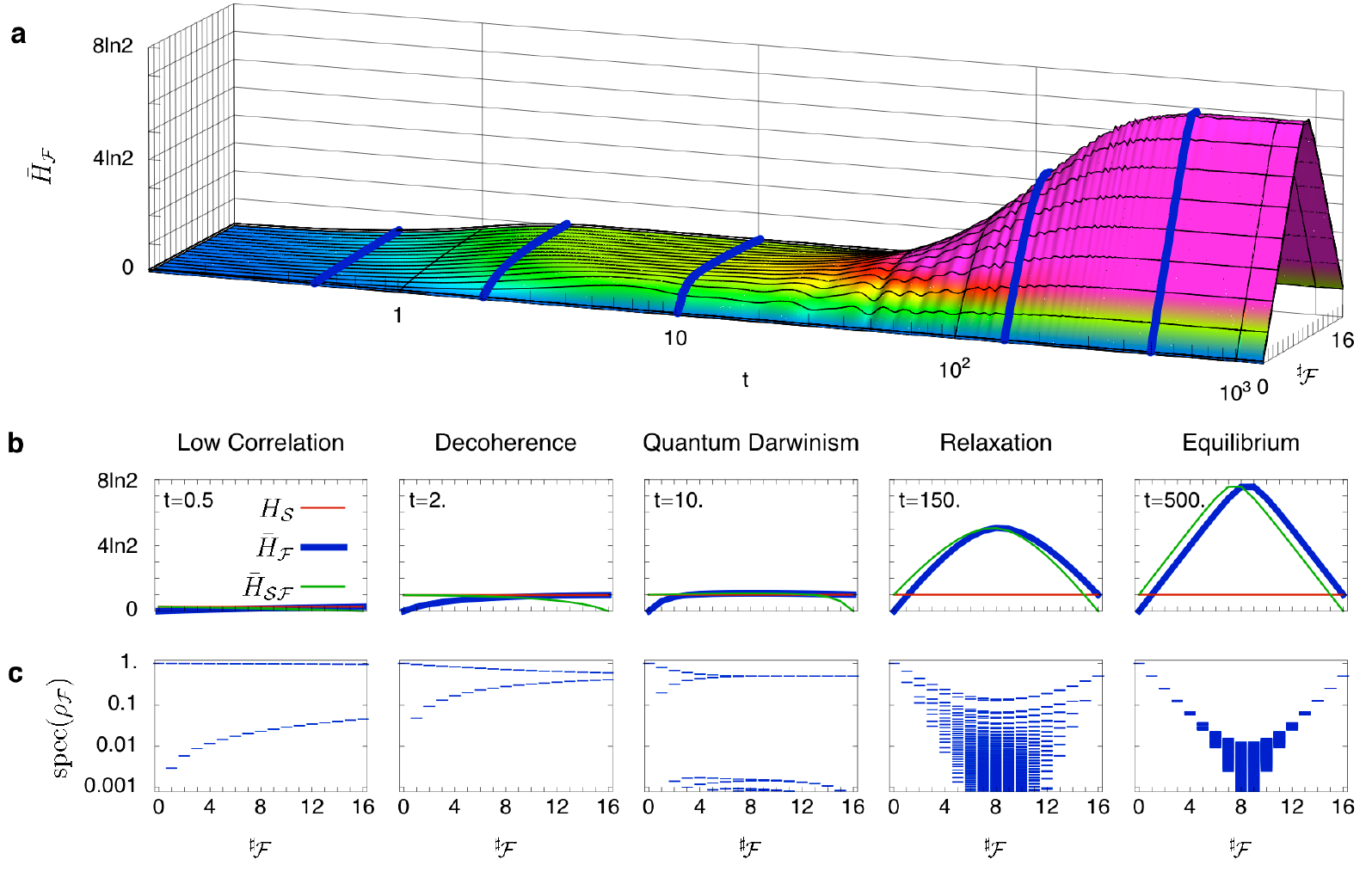}
  \caption{(Caption next page.)}
\end{figure}
\addtocounter{figure}{-1}
\begin{figure} [t!]
  \caption{(Previous page.)We study the spin universe described in figure \ref{diagram}.
(a) The average entropy $\bar{H}_\Frag$ of the fragment state $\rho_\Frag$ versus $\sharpof{\Frag}$ is the essential component of the mutual information, $\bar{\MI} = H_\Sys + \bar{H}_\Frag - \bar{H}_{\Sys \Frag}$, for understanding the rise and fall of redundancy and how it relates to branches in the global state of $\Sys \Env$.  Five time slices are are marked with thick lines.
(b) The three components of the mutual information ($H_\Sys$, $\bar{H}_\Frag$, and $\bar{H}_{\Sys \Frag}$) for each time slice.  Note that $\bar{H}_\Frag (\sharpof{\Frag}) = \bar{H}_{\Sys \Frag} (N-\sharpof{\Frag} )$ by the Schmidt decomposition and that---for all times after $\Sys$ is initially decohered---$H_\Sys$ is essentially equal to $H_\Sys^{\mathrm{max}} = \ln 2$.
(c) The eigenvalues of the state $\rho_\Frag$ (which determine $H_\Frag$) for the same five time slices.  The $n$-th largest value plotted is the average of the $n$-th eigenvalue of each choice of $\Frag$.
\textbf{Low correlation} ($t = 0.5$) exists before there are significant interactions, and there is just a single dominant branch corresponding to the initial product state. 
\textbf{Decoherence} ($t = 2$) produces two branches in the global state, one for each of the pointer states $\ket{\uparrow}$ and $\ket{\downarrow}$ of $\Sys$.  The system has been decohered by the environment at this point, but only very large fragments are fully correlated with $\Sys$.  Observers measuring less than half of the environment will \emph{not} be able to deduce the state of the system; as yet, there is no objectivity.  The widely separated eigenvalues for $\sharpof{\Frag} < N/2$ imply that the global branch structure is not accessible to local observers.
\textbf{Quantum Darwinism} ($t = 10$) is characterized by the fact that even small fragments $\Frag$ reveal the state of $\Sys$---and hence which branch the observer is on.   Since $\Frag$ only interacts with $\Sys$, there can only be two branches and the entropy $\bar{H}_\Frag$ is bounded by $H_\Sys^{\mathrm{max}} = \ln 2$.  By symmetry, the same is true for $\bar{H}_{\Sys \Frag}$. The tiny eigenvalues rising from below are the early indications of mixing.
\textbf{Relaxation} ($t = 150$) causes the number of significant eigenvalues to expand because $\Frag$ now entangles with its complement $\FragBar =  \bigotimes_{i \notin F} \Env_i$ in addition to $\Sys$.  The two branches (corresponding to the two dominant eigenvalues of $\rho_\Frag$) are beginning to divide, so that knowing a small fragment of the original branch no longer suffices to specify the pointer state of its root.  The entropies $\bar{H}_\Frag$ and $\bar{H}_{\Sys \Frag}$ quickly exceed $H_\Sys^{\mathrm{max}} = \ln 2$. 
\textbf{Equilibrium} ($t = 500$) follows. The state $\rho_\Frag$ approaches the maximally mixed matrix for $\sharpof{\Frag} < N/2$, so the eigenvalues of $\rho_\Frag$ are clustered around $1/\dim (\Frag) = 2^{-\sharpof{\Frag}}$.   The entropy $\bar{H}_\Frag$ approximately saturates its maximum, \eqref{frag-entropy-bound}.  The global branch structure is destroyed and the composite system $\Sys \Env$ cannot be given a classical description.
}
   \label{entropies}
\end{figure}


\begin{figure} [bt!] 
  \centering 
  \includegraphics[width=\medfactor\columnwidth]{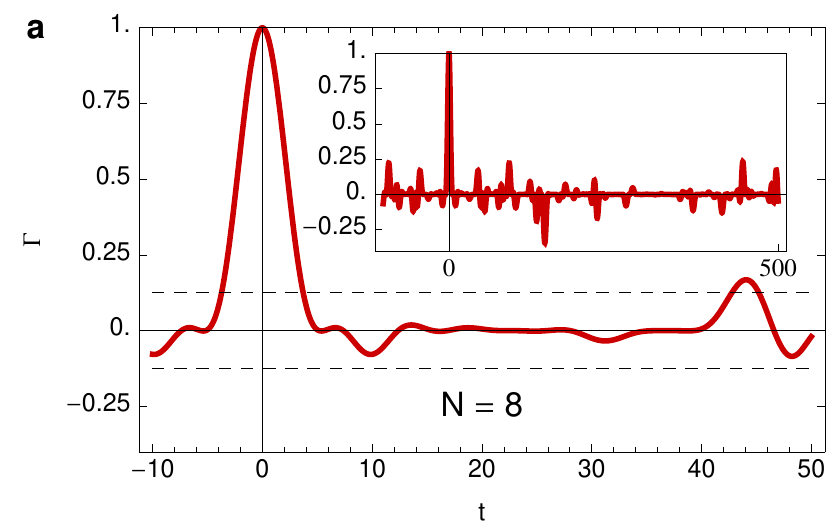} 
  \includegraphics[width=\medfactor\columnwidth]{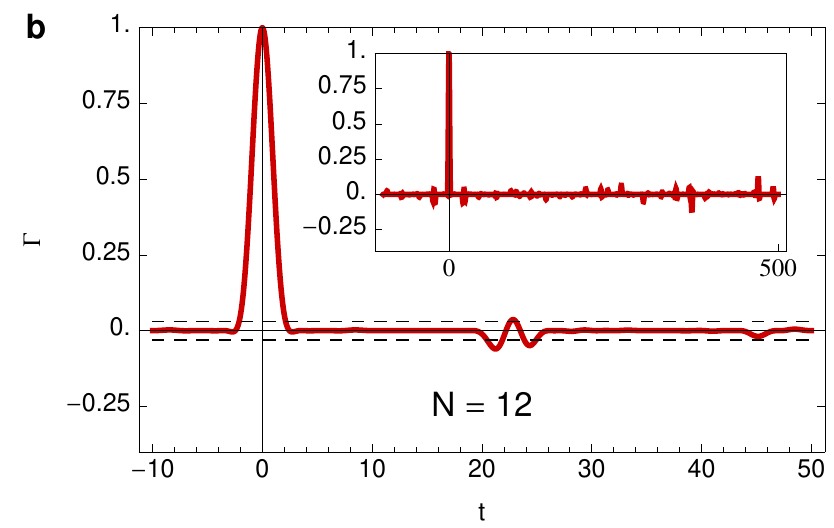}
  \caption{The decoherence factor $\Gamma$, \eqref{gamma}, for coupling constants $d_i$ chosen from a centered normal distribution with standard deviation $\sigma_d = 0.1$.  $\Gamma$ can be expressed as a product of cosines that are statistically independent for times $t \gg \sigma_d^{-1}$.  The statistical behavior of $\Gamma$ is seen in this long-time plot, and it can be shown that fluctuations away from zero are exponentially suppressed in the thermodynamic limit.  The dashed lines are at $\Gamma = \pm 2 \sigma_\Gamma$, with  $ \sigma_\Gamma^2 = 2^{-N}$ \eqref{sigma-gamma}.  (a) $N=8$.   (b) $N=12$.  The insets shows a longer time scale. Fluctuations for $N = 16$ would be too small to plot.}
  \label{decoh-factor}
\end{figure}

\section{Dynamics}

But of course, we know that observers can find out quite a bit about a system by interacting with much less than half of its environment.  This is because decoherence is ubiquitous in nature \cite{Zeh1970, Zurek2003a, JoosText, SchlosshauerText} and redundancy is produced universally by decoherence in the absence of coupling between different parts of the environment \cite{Zwolak00}.  However, realistic environments can have significant interactions between parts, so it's important to study these interactions and their effect on redundancy.  To see how high-redundancy states form through decoherence and how they can relax to a typical non-redundant state, we consider a model of a single qubit $\Sys$ ($D_\Sys =2$) monitored by an environment $\Env = \bigotimes_{i=1}^N \Env_i$ of $N$ spins ($D_\Env =2$)
\begin{eqnarray}
\Glob = \Sys \otimes \Env \cong \mathbb{C}^2 \otimes ( \mathbb{C}^2 )^{\otimes N}
\end{eqnarray}
with Hamiltonian
\begin{eqnarray}
\label{hamiltonian}
{\bm \hat{H}} = {\bm \hat{\sigma}}_\Sys^z  \otimes \sum_i d_i {\bm \hat{\sigma}}^z_i + \sum_{j,k} m_{j k}  {\bm \hat{\sigma}}^z_j \otimes  {\bm \hat{\sigma}}^z_k ,
\end{eqnarray}
where $d_i$ are the system-environment couplings and $m_{i j}$ are the intra-environment couplings.  We take the initial state to be 
\begin{eqnarray}
\label{initial-state}
\ket{\Psi_0} = \frac{1}{\sqrt{2^{N+1}}} \left[ \ket{\uparrow} + \ket{\downarrow} \right] \otimes \left[ \ket{+1} + \ket{-1} \right]^{\otimes N} .
\end{eqnarray}
For clarity, we denote the states of $\Sys$ with arrows ($\ket{\uparrow}, \ket{\downarrow}$) and the states of the $\Env_i$ with signs ($\ket{+1}, \ket{-1}$).  (There are several ways to relax this model for greater generality, but they are unnecessary for elucidating the key ideas.  We discuss generalizations at the end of this article.)


We use a numerical simulation ($N = 16$) to illustrate the build up of redundancy from the initial product state, and the subsequent transition to a typical non-redundant state.   (See figure \ref{diagram}.)  The couplings are selected from a normal distribution of zero mean and respective standard deviations $\sigma_d$ and $\sigma_m$.  Our key assumption to produce a high-redundancy state will be that $\Sys$ is coupled to the $\Env_i$ more strongly than the $\Env_i$ are coupled to each other ($\sigma_d \gg \sigma_m$).  This is an excellent approximations for many environments (e.g. a photon bath \cite{Riedel2010, Riedel2011}, where effectively $\sigma_m=0$) but not all (e.g. a gas of frequently colliding molecules).  This is the only condition that physically selects $\Sys$ as distinguished from the $\Env_i$.  For brevity, we'll call the timeframe $t \ll \sigma_m^{-1}$ the \emph{pure decoherence} regime and $t \not\ll \sigma_m^{-1}$ the (intra-environmental) \emph{mixing} regime. (We have set $\hbar = 1$.  In this article, we refer to interactions between spins within the environment as ``mixing''.)  

In addition to the two timescales $\sigma_d^{-1}$ and $\sigma_m^{-1}$ set by the typical size of the interaction terms, we also are interested in the times $\tau_d \equiv (\sqrt{N} \sigma_d)^{-1}$ and $\tau_m \equiv (\sqrt{N} \sigma_m)^{-1}$ which scale with the size of the environment.  Roughly, $t \gtrsim \sigma_d^{-1}$ and $t \gtrsim \sigma_m^{-1}$ are times for which the actions associated with couplings between individual spins (including the system qubit) are appreciable.  The earlier periods $t \gtrsim \tau_d$ and $t \gtrsim \tau_m$ are the times for which the collective action of the $N$ environment spins (on the the system and the environment itself, respectively) is strong.

Figure \ref{spins-PIP} shows the rise and fall of redundancy in the environment for our model, as well as the quantum mutual information between $\Sys$ and $\Frag$ as a function of fragment size $\sharpof{\Frag}$.  The maximum entropy of $\Sys$ is one bit: $H_\Sys^{\mathrm{max}} = \ln 2$.  The system is decohered, $H_\Sys \approx H_\Sys^{\mathrm{max}}$, when the environment becomes fully entangled with it, $\bar{\MI}(f = 1) \approx 2 H_\Sys^{\mathrm{max}}$, and this holds after $t \sim \tau_d$.  However, the mutual information does not form a plateau indicative of redundancy until $t \sim \sigma_d^{-1}$.  The plateau at $\bar{\MI} \approx H_\Sys^{\mathrm{max}}$ corresponds to approximately complete classical information about $\Sys$ available in most fragments $\Frag$ for $\sharpof{\Frag}$ not near $0$ or $N$.  But once enough time passes for the mixing to become significant, $t \sim \tau_m$, this structure is destroyed and the plot takes the form characteristic of typical non-redundant states.

To better illustrate what is going on, the average entropy $\bar{H}_{\Frag} (f)$ is plotted in figure \ref{entropies}a. During pure decoherence, $\bar{H}_{\Frag}$ saturates at $H_\Sys^{\mathrm{max}}$ for $\sharpof{\Frag}$ away from $0$ and $N$.  However, once the mixing during relaxation becomes substantial, $\bar{H}_{\Frag}$  approaches its maximum values consistent with the dimensionality of $\Frag$ and the symmetry ($H_{\Frag} = H_{\Sys \FragBar}$ where $\FragBar =  \bigotimes_{i \notin F} \Env_i$) imposed by the Schmidt decomposition:
\begin{eqnarray}
\label{frag-entropy-bound}
H_{\Frag}^{\mathrm{max}} &= \mathrm{Min}[\sharpof{\Frag}, N+1-\sharpof{\Frag}]  \ln 2, \\
\label{sys-frag-entropy-bound}
H_{\Sys \Frag}^{\mathrm{max}} &= \mathrm{Min}[\sharpof{\Frag} +1, N-\sharpof{\Frag}] \ln 2 .
\end{eqnarray}
In figure \ref{entropies}c, the eigenvalues for the corresponding state $\rho_\Frag$ are likewise plotted in both regimes.   This shows the formation and destruction of branches characteristic of quantum Darwinism \cite{Zurek2000, Zurek2009, Ollivier2004, Blume-Kohout2005}, and is suggestive of Everett's relative states \cite{EverettPhD, Everett1957}.  For pure decoherence, there are two dominant eigenvalues, corresponding to the entropy $H_\Frag$ capped at $H_\Sys^{\mathrm{max}}$.  As the mixing becomes important, the number of significant eigenvalues of $\rho_\Frag$ quickly rises and pushes the entropy to its maximum.  

\section{Branching}

We can develop a good intuition for this behavior by considering branches in the global state \cite{Zeh1970, Paz2009} of $\Sys \Env$.  Suppose that at a given moment the state can be decomposed as 
\begin{eqnarray}
\ket{\psi} = \sum_{q = 1}^{Q} \gamma_q \ket{\psi_q} = \sum_{q = 1}^{Q} \gamma_q \ket{s_q} \ket{e_q^{(1)}} \cdots  \ket{e_q^{(N)}}
\end{eqnarray}
for some small number $Q$ of orthogonal product state branches $\ket{\psi_q}$.  For $t=0$, we can have $Q=1$ since the initial state is a product state. In the decoherence regime (with approximate equality) we can have $Q=2$, i.e. a generalized GHZ state \cite{GHZstate}.  But once the environment begins to mix, $Q \gg 1$.  This gives a way for understanding the proliferation of eigenvalues in $H_\Frag$.  For any choice of fragment $\Frag$, its entropy $H_\Frag$ is bounded from above both by $H_\Frag^{\mathrm{max}}$ \eqref{frag-entropy-bound} \emph{and} by the entropy of the branch weights $\abs{\gamma_q}^2$, because the Schmidt decomposition associated with the cut $\Frag$-$\Sys \FragBar$ cannot have any more than $Q$ branches.  (See figure \ref{entropies}.)  More precisely, the spectrum of the fragment state $\rho_\Frag$ cannot be more mixed than the probability distribution $\abs{\gamma_q}^2$ according to the majorization partial order \cite{MarshallText, AlbertiText} for any choice of $\Frag$.

With this intuition in hand, we now derive the behavior seen in our model in the next two sections for large $N$; mathematical details can be found in the Appendix.



\section{Pure decoherence}

In the pure decoherence regime, $t \ll \sigma_m^{-1}$, both decoherence \cite{Zurek1981, Zurek1982} and quantum Darwinism \cite{Ollivier2004, Blume-Kohout2008, Zwolak2009, Zwolak2010} are well understood (even with $D_\Sys > 2$). 
The single decoherence factor of the two-state system quantifies the suppression of the off-diagonal terms of the density matrix $\rho^\Sys_t$ with time:
\begin{eqnarray} 
\label{rho-sys}
\rho^\Sys (t) = \frac{1}{2} \left( \begin{array}{cc} 1 & \Gamma(t)  \\ \Gamma (t) & 1 \end{array} \right),\\
\label{gamma}
\Gamma = \prod_{i=1}^N \cos (2 d_i t).
\end{eqnarray} 
The entropy of the two-dimensional state $\rho_\Sys$ \eqref{rho-sys} is then
\begin{eqnarray}
H_\Sys &=  \ln 2 - \Gamma \arctanh \Gamma - \ln \sqrt{1 - \Gamma^2} \\
\label{h-s-approx}
&\approx  \ln 2 -  \frac{1}{2} \Gamma^2 ,
\end{eqnarray}
where the approximation is valid for small $\Gamma$.  The average mutual information between $\Sys$ and $\Frag$ is
\begin{eqnarray}
\bar{\MI}(f) &= H_{\Sys} + \bar{H}_{\Frag}(f) - \bar{H}_{\Sys \Frag}(f) \\
\label{MI-pure-decoherence}
&\approx  \ln 2 -  \frac{1}{2}  (\Gamma^2 +\Gamma_\Frag^2 - \Gamma_\FragBar^2) ,
\end{eqnarray}
where $\Gamma_\Frag = \prod_{i \in F} \cos (2 d_i t)$, and $\Gamma_\FragBar = \prod_{i \notin F} \cos (2 d_i t)$.

The short and long time limits are illuminating.  For $t \ll \sigma_d^{-1}$ and large $N$, $\cos (2 d_i t) \approx 1-2 d_i^2 t^2$ so 
\begin{eqnarray}
\label{gamma-exp}
\Gamma \approx  \exp(-2 t^2 \sum_i d_i^2) \sim  \exp(-2 t^2 \sigma_d^2 N) .
\end{eqnarray}
Therefore, the system is essentially decohered when $t \sim \tau_d$, and $R_\delta \sim 1$.  The ensuing period $\tau_d \lesssim t \lesssim \sigma_d^{-1}$ exhibits quantum Darwinism. The system remains decohered but each spin in the environment continuously collects more and more information about the system.  Consequently, the redundancy steadily rises because the number of spins that must be measured by an observer to determine the state of the system falls.  This continues until $t \sim \sigma_d^{-1}$, when the phases associated with the action of the $\Env_i$ on $\Sys$ are of order unity.  At this point, the classical plateau of the mutual information congeals and $R_\delta \sim N$.

We can be precise by looking at $t \gg \sigma_d^{-1}$, when the values of the cosines on the rhs of \eqref{gamma} will act as \emph{independent} random variables \cite{Wintner1933}.  The statistical behavior is described by the time-averaged expectation values
\begin{eqnarray} 
\aavg{\Gamma} = 0, \\ 
\label{sigma-gamma} 
\sigma_\Gamma^2 \equiv \aavg{\Abs{\Gamma}^2} = 2^{-N},
\end{eqnarray}
since $\aavg{\cos^2 (2 d_i t)} = 1/2$.  

In other words, the decoherence factor $\Gamma$ has a Gaussian fall from unity for short times, and fluctuates around zero thereafter. This is illustrated in figure \ref{decoh-factor}.  The fluctuation of $\Gamma$ away from zero are exponentially suppressed, so fluctuations of $H_\Sys$ away from $H_\Sys^{\mathrm{max}} = \ln 2$ are similarly tiny. $\Gamma_\Frag$ and $\Gamma_\FragBar$ have the same behavior [with the respective replacements $N \to fN$ and $N \to (1-f) N$] so 
\begin{eqnarray}
\bar{\MI}(f) &\approx  \ln 2 -  \frac{1}{2}  \left[ 2^{-N} + 2^{- f N} - 2^{-(1-f)N} \right] \\
&\to \ln 2
\end{eqnarray} 
for $f \neq 0,1$ in the thermodynamic limit. This forms the robust classical plateau at $\bar{\MI} = \ln 2 = H_\Sys^{\mathrm{max}}$.

Although we concentrate here on the large time limit $t \gg \sigma_d^{-1}$ for the sake of rigor, note that the plateau starts forming at $t \sim \tau_d$ and finishes at $t \sim \sigma_d^{-1}$.  Indeed, even weak interactions lead to reliable redundancy \cite{Zwolak00}, a result that holds for higher dimensional subsystems. In particular, the ubiquitous real-life case of collisional decoherence through scattered light \cite{Riedel2010, Riedel2011} demonstrates how many weak correlations add up to huge redundancies.


\section{Mixing within the environment}

In the mixing regime, $t \not\ll \sigma_m^{-1}$, interactions within the environment force distinct records about $\Sys$ stored in the $\Env_i$ to intermingle, making it more difficult on average to determine the state of $\Sys$ by sampling a given fragment $\Frag$.  For large times, the mutual information between $\Sys$ and a typical $\Frag$ is nearly zero unless $f \ge \nicefrac{1}{2}$, i.e. an observer is unable to tell anything at all about the system until he makes a measurement on almost \emph{half} the environment.  Although the same amount of entanglement and information exists between $\Sys$ and $\Env$ regardless of mixing within the environment, the mixing spreads this information globally, rendering it locally inaccessible.  Information about $\Sys$ is no longer confined to the subsystems of $\Env$, but is stored in the correlations between them.  Similarly, one learns \emph{nothing} about whether or not a pair of playing cards are the same suit by looking at just one card.

To see this analytically, we now show that $\rho_\Frag$ will tend to the maximally mixed state $\rho_\Frag^\infty = I/2^{f N}$ for large times.  First, note that $\rho_\Frag$ agrees with $\rho_\Frag^\infty$ on the diagonal in the $z$-basis $\ket{\vec{r}}$, where $\ket{\vec{r}} = \bigotimes_{j \in F}  \ket{r_j}$, $r_j = \pm 1$, is a state of $\Frag$ specified by the vector $\vec{r}$.  The off-diagonal elements of $\rho_\Frag$ are suppressed by the factors
\begin{eqnarray}
\label{delta-g}
\Delta_{(\vec{r} - \vec{r}')} = \prod_{k \notin F} \left[ \cos \left(t \sum_{j \in F} (m_{j k} + m_{k j}) (r_j - r_j ' ) \right) \right],
\end{eqnarray}
which are analogous to $\Gamma$.  For $t \gg \sigma_m^{-1}$, the cosines will act like independent random variables and tend to cancel.  To be specific, $\aavg{\abs{\Delta_{(\vec{r} - \vec{r}')}}^2} = 2^{-(1-f) N}$ for $\vec{r} \neq \vec{r}'$.

For large times, one can show that the chance of an exponentially small fluctuation in $\rho_\Frag$ away from the maximally mixed state becomes exponentially unlikely in the thermodynamic limit:
\begin{eqnarray}
\label{trace-norm-fluctuations}
\mathrm{Pr}[T(\rho_\Frag, \rho_\Frag^\infty) > e^{- \kappa N}] \le e^{- 2 \kappa N} ,
\end{eqnarray}
where $T = T(\rho_\Frag, \rho_\Frag^\infty) = ||\rho_\Frag - \rho_\Frag^\infty||_1/2$ is the trace distance and $\kappa$ is a strictly positive constant for $f < \nicefrac{1}{2}$.  It is in this sense that $\rho_\Frag$ approaches the maximally mixed state for $f < \nicefrac{1}{2}$.  The Fannes-Audenaert inequality \cite{Fannes1973, Audenaert2007} then implies that exponentially tiny fluctuations in $H_\Frag$ are likewise exponentially unlikely over large times.  In that sense we say that
\begin{eqnarray}
H_\Frag \to H[\rho_\Frag^\infty] = f N \ln 2
\end{eqnarray}
as $N \to \infty$ for all $\Frag$ with $f < \nicefrac{1}{2}$.
With only a minor modification, the same argument can be applied to $\rho_{\Sys \Frag}$ to show $H_{\Sys \Frag} \to (f N + 1) \ln 2$.  We know from \eqref{h-s-approx} that $H_{\Sys} \to \ln 2$, so 
\begin{eqnarray}
\MI_{\Sys : \Frag} = H_{\Sys} + H_{\Frag} - H_{\Sys \Frag} \to 0
\end{eqnarray}
for all fragments satisfying $f < \nicefrac{1}{2}$.  Since $H_{\Env} = H_{\Sys} \to \ln 2$, we get $\bar{\MI}(f=1) \to 2 \ln 2$, and so by the anti-symmetry we know $\bar{\MI}(f) \to 2 \ln 2$ for $f > \nicefrac{1}{2}$.  

This explains the persistent step-function shape of the mutual information for large times as plotted in figure \ref{spins-PIP}.  This is the same form of the mutual information obtained with overwhelming probability by a state selected randomly from the global Hilbert space.


\section{Discussion}

Decoherence \cite{SchlosshauerText, JoosText, Zurek2003a} is crucial for understanding how classicality can arise in a purely quantum universe.  However, concentrating on individual systems (even while accounting for their interaction with the environment) leaves much to understand about global states.  

Quantum Darwinism has sharpened the vague idea that, based on our everyday observation of the effectiveness of our classical view of the world, there must be something very special about quasi-classical global states.  Hilbert space is dominated by non-redundant states, and these are not consistent with the high redundancy observers take for granted when they extrapolate an independent reality based on local interactions with the immediate environment.  

Quantum Darwinism shows how high redundancy can arise from decoherence. However, in many-body systems branching states with large redundancy cannot last forever.  The average mutual information $\bar{\MI}(f)$ approximates a step-function for almost all the states in Hilbert space,  so sampling ergodically produces such states with near certainty.  Therefore, relaxing to equilibrium necessarily means the destruction of redundancy.

If desired, our model can be generalized.  ``Unbalanced'' initial states of the system \cite{Riedel2011} or of the environment \cite{Zwolak2010}, such as $(2 \ket{\uparrow} + i \ket{\downarrow})/\sqrt{5}$, do not change the qualitative results.  The mutual information plateau will form lower at $H_\Sys^{\mathrm{max}} < \ln 2$ to agree with the maximum entropy of the system, and the limiting state $\rho_\Frag^\infty$ will change, but the factors $\Delta_{(\vec{r}-\vec{r}')}$ controlling fluctuations in $\rho_\Frag$ away from $\rho_\Frag^\infty$ will still be exponentially suppressed. The general unbalanced case is handled in the Appendix.  

We emphasize that the commuting nature of the interactions is very natural; the interaction terms between macroscopic objects (scattering) are almost always diagonal in position, a fact that can be traced back to real-world Hamiltonians.  Adding a self-Hamiltonian for $\Sys$ or the $\Env_i$ diagonal in the $z$-basis will not change any of our information theoretic results, since all the relevant density matrix spectra will be the same.   Self-Hamiltonians for $\Sys$ that do not commute with \eqref{hamiltonian} partially inhibit decoherence itself \cite{Zurek1993a, Zurek2003a, Cucchietti2005}, but will not stop the information mixing in the environment.  In general, system self-Hamiltonians that do not commute with the system-environment interaction are necessary to produce the repeated branching that occurs in nature. For example, the rate of diffusion for the quantum random walk of an object decohered by collisions with a gas is set by the size of the self-Hamiltonian $p^2/2m$ relative to the strength of the scattering \cite{Diosi2000}. An enticing subject for future research would be the analysis of quantum Darwinism in the case of repeated branchings due to a non-commuting self-Hamiltonian, and the dependency of redundancy on the rate of branching.  In particular, we expect strong connections \cite{Dalvit2001, Dziarmaga2004, Paz1993} with the quantum trajectories \cite{CarmichaelText} and consistent histories \cite{GriffithsText, Gell-Mann1990a} formalisms.

Our simple model has highlighted how important the relative strengths of couplings are for the distinction between system and environment, and the development of redundancy.  Indeed, coupling strength is the only thing here that distinguished the system from the environment.  If we had not assumed that the mixing within the environment was slower than the decoherence of the system, there would be no intermediate timespan $\sigma_d^{-1}  \ll t \ll \sigma_m^{-1}$ and the mixing would destroy redundancy before it had a chance to develop.  

Such mixing would seem unimportant when studying the decoherence of a system of a-priori importance, but it's illuminating for understanding what distinguishes certain degrees of freedom in nature as preferred.  A large molecule localized through collisional decoherence by photons is immersed in an environment with insignificant mixing \cite{Joos1985}, and so is recorded redundantly \cite{Riedel2010, Riedel2011}, but a lone argon atom in a dense nitrogen gas is not. Whether an essentially unique quasi-classical realm \cite{Gell-Mann2007, Hartle2011} can be identified from such principles is a deep, open question \cite{Kent1996,Dowker1996} about the quantum-classical transition.  


\ack
We thank Robin Blume-Kohout, Eric Konieczny, Greg Kuperberg, David Moews, and Jon Yard for discussion.  This research is supported by the U.S. Department of Energy through the LANL/LDRD program and by the John Templeton Foundation.



\appendix

\section*{Appendix}
\setcounter{section}{1} 

Here we discuss decoherence factors in the thermodynamic ($N \to \infty$) and large time ($t \to \infty$) limit.   Recall that our model consists of a single qubit $\Sys$ monitored by an environment $\Env = \bigotimes_{i=1}^N \Env_i$ of $N$ spins 
\begin{eqnarray}
\label{hilbert-space-supp}
\Glob = \Sys \otimes \Env \cong \mathbb{C}^2 \otimes \left( \mathbb{C}^2 \right)^{\otimes N}
\end{eqnarray}
with Hamiltonian
\begin{eqnarray}
\label{hamiltonian-supp}
\mathbf{\hat{H}} = {\bm \hat{\sigma}}_\Sys^z \otimes \sum_i d_i {\bm \hat{\sigma}}^z_i + \sum_{j,k} m_{j k}  {\bm \hat{\sigma}}^z_j \otimes  {\bm \hat{\sigma}}^z_k ,
\end{eqnarray}
where $d_i$ are the system-environment couplings and $m_{i j}$ and the environment-environment couplings.  The initial state is
\begin{eqnarray}
\label{initial-state-supp}
\ket{\Psi_0} =  \left[ \beta_\uparrow \ket{\uparrow} + \beta_\downarrow \ket{\downarrow} \right] \otimes \left[ \alpha_{+1} \ket{+1} + \alpha_{-1} \ket{-1} \right]^{\otimes N}
\end{eqnarray}
where $ \Abs{\beta_\uparrow}^2+\Abs{\beta_\downarrow}^2  = \Abs{\alpha_{+1}}^2+\Abs{\alpha_{-1}}^2 = 1$.  Let $\Frag =  \bigotimes_{i \in F} \Env_i$ be a fragment of $\Env$, where $F \subset \{1, \ldots, N\}$ and $\sharpof{\Frag} = \Abs{F} = f N$,  $0 \le f \le 1$.  The complement fragment is $\FragBar =  \bigotimes_{i \notin F} \Env_i$.

Let us break up the evolution into commuting unitaries
\begin{eqnarray}
\mathbf{\hat{U}} = e^{- i \mathbf{ \hat{H}} t} = \mathbf{ \hat{U}}_{\Sys \Frag} \mathbf{ \hat{U}}_{\Sys \FragBar} \mathbf{ \hat{U}}_{\Frag \Frag} \mathbf{ \hat{U}}_{\Frag \FragBar} \mathbf{ \hat{U}}_{\FragBar \FragBar},
 \end{eqnarray}
 labeled by the subsystems they couple, e.g.
 \begin{eqnarray}
\mathbf{\hat{U}}_{\Frag \FragBar} = \exp\left[-i t \sum_{j \in F} \sum_{k \notin F} (m_{j k} + m_{k j}) {\bm \hat{\sigma}}^z_j \otimes {\bm \hat{\sigma}}^z_k \right].
 \end{eqnarray} 
The single decoherence factor of the two-state system quantifies the suppression of the off-diagonal terms of the density matrix $\rho^\Sys$ with time:
\begin{eqnarray}
\rho^\Sys (t) = \left( \begin{array}{cc} \abs{\beta_\uparrow}^2 & \beta_\uparrow \beta_\downarrow^* \Gamma(t)  \\  
\beta_\uparrow^* \beta_\downarrow\Gamma^* (t) &  \abs{\beta_\downarrow}^2 \end{array} \right),
\end{eqnarray}
where
\begin{eqnarray} 
\Gamma (t)  &= \prod_{i=1}^N  \left[ \abs{\alpha_{+1}}^2 e^{-2 i t d_i} + \abs{\alpha_{-1}}^2 e^{2 i t d_i} \right].
\end{eqnarray}
We are interested in the statistical behavior of this term for times large compared to the $d_i$, especially for large values of $N$.  For any function $\mu(t)$, we can define a random variable $Z$ over a rigorous probability space through the cumulative distribution function
\begin{eqnarray}
F_{Z}(z) \equiv P[Z > z] \equiv \lim_{T \to \infty} \frac{1}{T} \lambda \{ t \in [0,T] \vert \mu(t) > z\} ,
\end{eqnarray}
provided the limit exists.  (Here, $\lambda$ is the Lebesgue measure).  To be suggestive, we can denote expectation values over long times constructed with such a random variable using the time-dependent function:  $\aavg{\mu^2}$, $\aavg{\log(\mu)}$, etc. A result of the theory of almost periodic functions \cite{BohrText, ZhangText} is that random variables defined in this way from periodic functions of time are statistically independent if their periods are linearly independent over the rationals \cite{Wintner1933}.  Unless the $d_i$ are chosen to be exactly linearly dependent, this means that
\begin{eqnarray}
\aavg{\Gamma} &= 0,\\
\aavg{\abs{\Gamma}^2} &= \prod_{i=1}^N \aavg{\Abs{a \, e^{-2 i t d_i} + (1-a) e^{2 i t d_i} }^2} = \left[a^2 + (1-a)^2 \right]^N ,
\end{eqnarray}
since $\aavg{e^{-4 i t d_i}} =0$.  We have defined the probability $a = |\alpha_{+1}|^2$ and note that $\nicefrac{1}{2} \le a^2 + (1-a)^2 \le 1 $. 

Thus, so long as the environment isn't an exact eigenstate of the Hamiltonian ($a \neq 0,1$), fluctuations of the decoherence factor $\Gamma$ away from zero (as measured by the variance) are exponentially suppressed in the thermodynamic limit.  (For physical intuition about these results, see \cite{Zurek1982}.)  This sends 
\begin{eqnarray}
H_\Sys \to H_\Sys^{\mathrm{max}} = H_2[b], 
\end{eqnarray}
where $H_2[x]=-x \ln x -(1-x) \ln (1-x)$ is the binary entropy function and $b \equiv \abs{\beta_\uparrow}^2$.  Of course, $H_\Sys^{\mathrm{max}}  \le \ln 2$, with equality iff $b = \frac{1}{2}$. 

We can quickly extend this to a statement about quantum Darwinism in the case of pure decoherence ($m_{j k}$ negligible): 
\begin{eqnarray} 
\rho_\Frag &= \Tr_{\Sys \FragBar} \left[ \mathbf{\hat{U}}_{\Sys \Frag} \left( \rho^\Sys_0 \otimes \rho^\Env_0 \right)  \mathbf{\hat{U}}_{\Sys \Frag}^\dagger \right] \\
&= \abs{\beta_\uparrow}^2 \projector{\Frag \uparrow} + \abs{\beta_\downarrow}^2 \projector{\Frag \downarrow} \label{rho-frag-spectrum}
\end{eqnarray}
where $\ket{\Frag \uparrow} = \bigotimes_{j \in F} (\alpha_{+1} e^{-i d_j t}\ket{+1} + \alpha_{-1} e^{+i d_j t}\ket{-1} )$ is the pure state of $\Frag$ conditional on the system being up, and likewise for $\ket{\Frag \downarrow}$.  The decoherence factor of the rank-$2$ matrix $\rho_\Frag$ is $\Gamma_\Frag = \braket{\Frag \downarrow}{\Frag \uparrow}$ and 
\begin{eqnarray}
\Gamma_\Frag &=\prod_{j \in F}  \left[ a \, e^{-2 i t d_j} + (1-a) e^{2 i t d_j} \right] , \\
\aavg{\Gamma_\Frag} &= 0 , \\
\label{gamma-squared-avg}
\aavg{|\Gamma_\Frag|^2} &=\left[a^2 + (1-a)^2 \right]^{f N} .
\end{eqnarray}
For fixed $f > 0$, fluctuation in $\Gamma_\Frag$ will be exponentially suppressed in $N$, just like $\Gamma$.  With small $\Gamma_\Frag$,
\begin{eqnarray}
H_\Frag &= H_2 \left[ \frac{1}{2} \left(1+\sqrt{1-4b(1-b)(1-|\Gamma_\Frag|^2)} \right) \right] \\
& \approx H_\Sys^{\mathrm{max}}  - \chi(b) \frac{|\Gamma_\Frag|^2}{2}
\end{eqnarray}
where
\begin{eqnarray}
\chi(b) \equiv \left[\frac{4 b (1-b) \arctanh (1-2b)}{1-2b}\right]
\end{eqnarray}
and $0 < \chi(b) \le 1$ with $\chi(b) = 1$ iff  $b=\nicefrac{1}{2}$.  Therefore, $H_\Frag$ quickly approaches $H_\Sys^{\mathrm{max}}$. 
Further, because the global state is pure, we know $H_{\Sys \Frag} = H_{\FragBar}$ and so
\begin{eqnarray}
\MI_{\Sys : \Frag} &= H_{\Sys} + H_{\Frag} - H_{\Sys \Frag} \\
&\approx  H_\Sys^{\mathrm{max}} - \chi(b) \frac{\abs{\Gamma}^2 +\abs{\Gamma_\Frag}^2 - \abs{\Gamma_\FragBar}^2}{2}  .
\end{eqnarray}
For $f$ away from 0 and 1, all three decoherence factors are exponentially suppressed in $N$. This is the origin of the robust plateau on the plot of average mutual information.  One can show \cite{Riedel2011} this means the redundancy grows linearly with $N$.

Now we will extend this result to determine the statistical behavior of $H_\Frag$ and $H_{\Sys \Frag}$ when the interactions within the environment are not negligible.  This will let us show that for times large compared the couplings $m_{i j}$ the states of $\Frag$ and $\Sys \Frag$ become maximally mixed subject to constraints of the initial conditions.  First, under the evolution of $H$, the state $\rho_\Frag$ of the fragment is unitarily equivalent to 
\begin{eqnarray}
\tilde{\rho}_\Frag \equiv \Tr_{\Sys \FragBar} \left[ \mathbf{\hat{U}}_{\Sys \Frag} \mathbf{\hat{U}}_{\Frag \FragBar} \projector{\Psi_0} \mathbf{\hat{U}}_{\Frag \FragBar}^\dagger \mathbf{\hat{U}}_{\Sys \Frag}^\dagger  \right].
\end{eqnarray}
A bit of algebra gives
\begin{eqnarray}
\fl \matrixelement{\vec{r}}{\tilde{\rho}_\Frag}{\vec{r}'} =  \left[ b \,  e^{-i t \sum_{j \in F} d_j (r_j - r_j ' )} + (1-b)  e^{+i t \sum_{j \in F} d_j (r_j - r_j ' )} \right] \Delta_{(\vec{r} - \vec{r}')} \prod_{j \in F} \alpha_{r_j} \alpha_{r_j'}^*
\end{eqnarray}
where $\ket{\vec{r}} = \bigotimes_{j \in F} \ket{r_j}$, $r_j = \pm 1$, is a state of $\Frag$ specified by the vector $\vec{r}$.  
Above, 
\begin{eqnarray}
\label{delta-g} 
\fl  \Delta_{(\vec{r} - \vec{r}')} =  \prod_{k \notin F} \Bigg[ a  \exp  &\left( - i t \sum_{j \in F}  (m_{j k} + m_{k j}) (r_j - r_j ' ) \right) \\
&+ (1-a) \exp \left( i t \sum_{j \in F} (m_{j k} + m_{k j}) (r_j - r_j ' ) \right) \Bigg].
\end{eqnarray}
We want to show that the entropy $H_\Frag$ of $\rho_\Frag$ approaches its maximum value $f N H_2[a]$ for $f < \nicefrac{1}{2}$ by bounding the difference between $\tilde{\rho}_\Frag$ and the limiting state $\tilde{\rho}_\Frag^\infty$:
\begin{eqnarray}
\tilde{\rho}_\Frag^\infty = \bigotimes_{i \in F} \left[ a \projector{+1} + (1-a) \projector{-1} \right].
\end{eqnarray}
First, we will assume the case of a balanced initial environmental state, $a = \nicefrac{1}{2}$.  The Hilbert-Schmidt norm of the difference is
\begin{eqnarray}
\label{hs-norm}
\abs{\abs{\tilde{\rho}_\Frag - \tilde{\rho}_\Frag^\infty}}_{\mathrm{HS}}^2 &= \sum_{\vec{r}} \sum_{\vec{r}'  \neq \vec{r}} \abs{\matrixelement{\vec{r}}{\tilde{\rho}_\Frag}{\vec{r}'}}^2  \\
&\le \frac{1}{4^{fN}} \sum_{\vec{r}} \sum_{\vec{r}'  \neq \vec{r}} \abs{\Delta_{(\vec{r} - \vec{r}')}}^2 .
\end{eqnarray}
Now, we want to bound fluctuations of $H_\Frag$ away from its limiting value $H_\Frag^\infty$, the entropy of $\tilde{\rho}_\Frag^\infty$.  To do this, we use Audenaert's optimal refinement \cite{Audenaert2007} of Fannes' inequality \cite{Fannes1973} governing the continuity of the von Neumann entropy.  For any two density matrices $\rho_1$ and $\rho_2$ with trace norm distance $T = T(\rho_1,\rho_2) =\abs{\abs{\rho_1 - \rho_2}}_1/2$, the difference in their entropies $\Delta H$ is bounded as
\begin{eqnarray}
\abs{\Delta H} \le T \ln(D -1) + H_2[T]
\end{eqnarray}
where $D$ is the dimension of the matrices.  We will also use the bound between the trace norm distance and the Hilbert-Schmidt norm for Hermitian matrices, $\abs{\abs{\rho}}_1 \le \sqrt{D} \abs{\abs{\rho}}_{\mathrm{HS}}$, to get
\begin{eqnarray}
\label{1-2-norm}
T \le \frac{1}{2} \sqrt{2^{f N}} \abs{\abs{\tilde{\rho}_\Frag - \tilde{\rho}_\Frag^\infty}}_{\mathrm{HS}}.
\end{eqnarray}
Now we consider the likelihood of fluctuations in $T$ bigger than an arbitrary $T_0$:
\begin{eqnarray}
\label{pt}
\fl P[T > T_0] &\le P \left[ \abs{\abs{\tilde{\rho}_\Frag - \tilde{\rho}_\Frag^\infty}}_{\mathrm{HS}} > \frac{2 T_0}{2^{f N /2}} \right] = P \left[ \abs{\abs{\tilde{\rho}_\Frag - \tilde{\rho}_\Frag^\infty}}_{\mathrm{HS}}^2  > \frac{4 T_0^2}{2^{f N }} \right] .
\end{eqnarray}
By the definition of an expectation value, we know
\begin{eqnarray}
\label{phs}
P \left[ \abs{\abs{\tilde{\rho}_\Frag - \tilde{\rho}_\Frag^\infty}}_{\mathrm{HS}}^2  > \frac{4 T_0^2}{2^{f N }} \right]   \frac{4 T_0^2}{2^{f N }} \le \aavg{\abs{\abs{\tilde{\rho}_\Frag - \tilde{\rho}_\Frag^\infty}}_{\mathrm{HS}}^2} .
\end{eqnarray}
And, from \eqref{hs-norm}, we know
\begin{eqnarray}
\aavg{\abs{\abs{\tilde{\rho}_\Frag - \tilde{\rho}_\Frag^\infty}}_{\mathrm{HS}}^2} &\le \frac{1}{4^{f N}}  \sum_{\vec{r}} \sum_{\vec{r}'  \neq \vec{r}} \aavg{\abs{\Delta_{(\vec{r} - \vec{r}')}}^2} \\
&= \frac{1}{4^{f N}}  \sum_{\vec{r}} \sum_{\vec{r}'  \neq \vec{r}} \frac{1}{2^{(1-f) N}} \\
\label{ahs}
&\le  \frac{1}{2^{(1-f) N}}.
\end{eqnarray}
We calculated  $\aavg{\abs{\Delta_{(\vec{r} - \vec{r}')}}^2}$ exactly as we did $\aavg{\Gamma}$.  If $\vec{r}  \neq \vec{r}'$, the the sums over $j$ in \eqref{delta-g} are non-empty and---assuming the $m_{j k}$ aren't specially chosen to be linearly dependent over the rationals---each $k$-indexed term in the product of \eqref{delta-g} is statistically independent.  

Combining \eqref{pt}, \eqref{phs}, and \eqref{ahs}, we find
\begin{eqnarray}
\label{prob-limit}
P[T > T_0] \le \frac{2^{(2f-1) N}}{4 T_0^2}. 
\end{eqnarray}
So if $f<\nicefrac{1}{2}$, we can choose $T_0 = 2^{-(1/2-f) N/2}$ so that both $T_0$ and $P[T > T_0]$ are suppressed:
\begin{eqnarray}
P[T > 2^{-(1/2-f) N/2}] \le 2^{ -(1/2-f) N} .
\end{eqnarray}
In other words, as we take the size of the environment $N$ to infinity, exponentially tiny fluctuations in the trace norm distance $T = T( \tilde{\rho}_\Frag, \tilde{\rho}_\Frag^\infty)$ become exponentially unlikely.  It is in this sense that we say 
\begin{eqnarray}
\rho_\Frag \to \tilde{\rho}_\Frag^{\infty}
\end{eqnarray}
up to unitary equivalence.  We can slightly relax the Fannes-Audenaert inequality to make it a little more transparent:
\begin{eqnarray}
\abs{\Delta H} &\le  T_0 \ln(D -1) + H_2[T_0] \\
&\le T_0 \left[ 1 + fN \ln2 + \ln \frac{1}{T_0} \right].
\end{eqnarray}
So likewise for the entropy $H_\Frag$, exponentially tiny fluctuations are exponentially unlikely for large $N$.  It is in this sense that we say 
\begin{eqnarray}
H_\Frag \to H_\Frag^{\infty} = f N \ln 2 
\end{eqnarray}
for $f < \nicefrac{1}{2}$.

With only a minor modification, the same argument can be applied to $\rho_{\Sys \Frag}$ to show
\begin{eqnarray}
H_{\Sys \Frag} \to H_{\Sys \Frag}^{\infty} = H_{\Frag}^{\infty}  + H_\Sys^{\mathrm{max}}.
\end{eqnarray}
We know from \eqref{h-s-approx} that $H_{\Sys} \to H_\Sys^{\mathrm{max}}$, so $\MI_{\Sys : \Frag} = H_{\Sys} + H_{\Frag} - H_{\Sys \Frag} \to 0$ for $f < \nicefrac{1}{2}$.  Since $H_{\Sys} = H_{\Env}$, we get $\MI_{\Sys : \Frag} \to 2 H_\Sys^{\mathrm{max}}$ for $f=1$, so by the anti-symmetry we know $\MI_{\Sys : \Frag} \to 2 H_\Sys^{\mathrm{max}}$ for $f > \nicefrac{1}{2}$.  This gives exactly the step-function-shaped curve of a typical non-redundant state, so that $R_\delta \approx 2$, independent of $\delta$.

\begin{figure*} [bt]
  \centering 
  \includegraphics[width=\medfactor\textwidth]{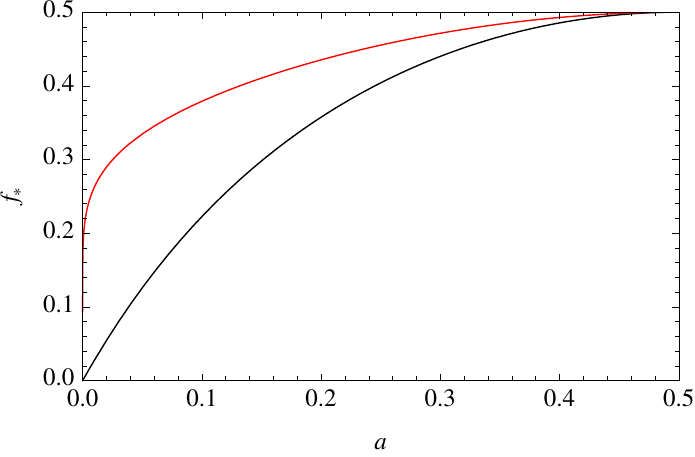}
  \caption{Our argument is valid for $f < f_*$, where $f_*$ is a function of how balanced the initials state of the environment is, as parameterized by $a = \abs{\alpha_{+1}}^2$.  Only $a \le \nicefrac{1}{2}$ is shown because $f_*(1-a) = f_*(a)$. The colors denote the value of $f_*$ without (black) and with (red) Schumacher compression.}
  \label{f-star}
\end{figure*}

If we directly extend this proof to the unbalanced case, $a \neq \nicefrac{1}{2}$, we make the replacement
\begin{eqnarray}
\frac{1}{2^{(1-f)N}} \longrightarrow \left[ a + (1-a)^2 \right]^{(1-f)N}
\end{eqnarray}
in \eqref{ahs}, but the factor of $2^{fN}$ from \eqref{1-2-norm} is unchanged.  This means we can show
\begin{eqnarray}
H_\Frag \to H_{\Frag}^{\infty} = fN H_2[a]
\end{eqnarray}
only for $f < f_*$ where
\begin{eqnarray}
f_* = \left[1 - \frac{1}{\log_2 [a^2 + (1-a)^2]} \right]^{-1}.
\end{eqnarray}
Note that $0 \le f_* \le \nicefrac{1}{2}$, and  $f_* = \nicefrac{1}{2}$ iff $a = \nicefrac{1}{2}$.  We can use Schumacher compression \cite{Jozsa1994, Schumacher1995} to improve $f_*$.  Take the typical sequence \cite{CoverText} of eigenvalues $\lambda_{\vec{r}} = \prod_{j \in F} \abs{\alpha_{r_j}}^2$ of $\tilde{\rho}_F^\infty$
\begin{eqnarray}
S_\delta \equiv \{ \vec{r} \mid  e^{-fN (H_2[a]+\delta)} \le \lambda_{\vec{r}} \le e^{-fN (H_2[a]-\delta)} \} ,
\end{eqnarray} 
and define the typical subspace $\Lambda_\delta \subset \Frag$, $\delta > 0$, as the subspace corresponding to those eigenvalues
\begin{eqnarray}
\Lambda_\delta \equiv \mathrm{span} \{ \ket{\lambda_{\vec{r}}} \mid \vec{r} \in S_\delta\}
\end{eqnarray} 
with projector $\Pi_\delta = \sum_{\vec{r} \in S_\delta} \projector{\vec{r}}$.  Then define $P_\delta \equiv \sum_{\vec{r} \in S_\delta} \lambda_{\vec{r}} = \Tr \Pi_\delta \tilde{\rho}_\Frag^\infty \Pi_\delta $ and the (normalized) density matrices
\begin{eqnarray}
\tilde{\eta}_\Frag \equiv \frac{1}{P_\delta} \Pi_\delta \tilde{\rho}_\Frag \Pi_\delta, \\
\tilde{\eta}_\Frag^{\infty} \equiv \frac{1}{P_\delta} \Pi_\delta \tilde{\rho}_\Frag^\infty \Pi_\delta .
\end{eqnarray} 
Use the triangle inequality to bound
\begin{eqnarray}
|| \tilde{\rho}_\Frag  - \tilde{\rho}_\Frag^{\infty}  ||_1 \le || \tilde{\rho}_\Frag  - \tilde{\eta}_\Frag  ||_1 + || \tilde{\eta}_\Frag  - \tilde{\eta}_\Frag^{\infty}  ||_1 + || \tilde{\eta}_\Frag^{\infty}  - \tilde{\rho}_\Frag^{\infty}  ||_1.
\end{eqnarray} 
The norms $|| \tilde{\rho}_\Frag  - \tilde{\eta}_\Frag  ||_1$ and $|| \tilde{\eta}_\Frag^{\infty}  - \tilde{\rho}_\Frag^{\infty}  ||_1$ can be handled with the close relationship between the fidelity and the trace distance: $1- F(\rho_1,\rho_2) \le D(\rho_1,\rho_2) \le \sqrt{ 1- F(\rho_1,\rho_2)^2}$. Using Hoeffding's inequality \cite{Hoeffding1963} we can show that these norms are suppressed exponentially in $N$.  Now we just bound $ || \tilde{\eta}_\Frag  - \tilde{\eta}_\Frag^{\infty}  ||_1$ using the Hilber-Schmidt norm where,  importantly, $\tilde{\eta}_\Frag$ and $\tilde{\eta}_\Frag^{\infty}$ live in a subspace of dimension $e^{f N H_2[a]}$ rather than $2^{f N}$.  This gives an improved range for the applicability of our argument: $f < f_*$ where
\begin{eqnarray}
f_* = \left[1 - \frac{H_2[a]}{\log_2 [a^2 + (1-a)^2]} \right]^{-1} .
\end{eqnarray}
The improvement on $f_*$ is depicted in figure A1.  This turns out to be the best we can do using the bound \eqref{gamma-squared-avg}.  It's possible to construct a counter-example matrix $\hat{\rho}_\Frag$ when $f > f_*$ that satisfies \eqref{gamma-squared-avg} but has limiting entropy $\hat{H}_{\Frag}^{\infty} = (1-f)N \ln[1/(a^2+(1-a)^2)] < H_{\Frag}^{\infty} $.

So, in the case that $a \neq \nicefrac{1}{2}$, we are only able to prove that $H_\Frag \to H_\Frag^{\infty} = f N \ln 2 $ and $\bar{\MI}(f) \to 0$ for $f < f_*$.  This means the redundancy can be bounded only by $R_\delta < 1/f_*$.  Now, $f_*$ is of order unity unless the initial state of the environmental spins are nearly eigenstates of the interaction Hamiltonian, so this is still a very strong upper bound on the redundancy.  In contract, $R_\delta$ grows linearly with $N$ for a branching state.


\bibliographystyle{ieeetr}
\bibliography{riedelbib}

\end{document}